\documentclass[journal,10pt]{IEEEtran}
\usepackage{cite}
\usepackage{bbding}
\usepackage{makecell}
\usepackage{graphicx}
\usepackage{float}
\usepackage{amsthm}
\usepackage[colorlinks=black,linkcolor=black,anchorcolor=black,citecolor=black]{hyperref}
\usepackage{amsmath,amssymb,amsfonts,bm,subfigure,lipsum,amsthm}
\usepackage{stfloats}
\usepackage{algorithm}
\usepackage{algorithmic}
\usepackage{mathrsfs}
\usepackage{graphicx}
\usepackage[table]{xcolor}
\usepackage{threeparttable}
\usepackage{textcomp}
\usepackage{xcolor}
\usepackage{color}
\usepackage{booktabs}
\usepackage{amsmath}
\usepackage{bbm}
\usepackage{multicol}
\usepackage{multirow}
\usepackage[english]{babel}
\usepackage{upgreek}
\usepackage{float}
\usepackage{colortbl}
\usepackage[table]{xcolor}
\usepackage{cuted}
\usepackage{epstopdf}
\newtheorem{remark}{Remark}
\newtheorem{theorem}{Theorem}

\def\BibTeX{{\rm B\kern-.05em{\sc i\kern-.025em b}\kern-.08em
    T\kern-.1667em\lower.7ex\hbox{E}\kern-.125emX}}
\usepackage{graphicx,graphics,color,epsfig,subfigure,graphpap,rotate}
\usepackage{times, verbatim, subfigure, epsfig, graphicx, latexsym}
\usepackage{url}
\usepackage{subfigure}

\definecolor{hellmagenta}{rgb}{1,0.75,0.9}

\definecolor{hellcyan}{rgb}{0.75,1,0.9}

\definecolor{hellgelb}{rgb}{1,1,0.8}

\definecolor{colKeys}{rgb}{0,0,1}

\definecolor{colIdentifier}{rgb}{0,0,0}

\definecolor{colComments}{rgb}{1,0,0}

\definecolor{colString}{rgb}{0,0.5,0}

\definecolor{darkyellow}{rgb}{1,0.9,0}

\begin{document}
\title{ Diffusion-enabled  Secure Semantic Communication  Against Eavesdropping }

\author{Boxiang~He,~Zihan~Chen,~Fanggang~Wang,~\IEEEmembership{Senior~Member,~IEEE},~Shilian~Wang,\\~Zhijin Qin,~\IEEEmembership{Senior~Member,~IEEE,}~and~Tony~Q.S.~Quek,~\IEEEmembership{Fellow,~IEEE}

\thanks{ Boxiang  He and Shilian Wang are with the
College of Electronic Science and Technology, National University of Defense Technology, Changsha 410073, China (e-mail: boxianghe1@bjtu.edu.cn;  wangsl@nudt.edu.cn).

Zihan Chen and Tony Q.S. Quek are with the  Information Systems Technology and
Design, Singapore University of Technology and Design, Singapore
(e-mail: zihan$\_$chen@mymail.sutd.edu.sg; tonyquek@sutd.edu.sg).

Fanggang Wang is with the School of Electronic and Information Engineering, Beijing Jiaotong University, Beijing 100044, China (e-mail:  wangfg@bjtu.edu.cn).

Zhijin Qin is with the Department of Electronic Engineering, Tsinghua
University, Beijing, 100084, China, and also with the Beijing National
Research Center for Information Science and Technology, Beijing, China, and
the State Key Laboratory of Space Network and Communications, Beijing,
China (e-mail: qinzhijin@tsinghua.edu.cn).

}

}

\maketitle

\begin{abstract}
This paper proposes a novel diffusion-enabled pluggable encryption/decryption modules design against semantic eavesdropping, where the pluggable modules are optionally assembled into the  semantic communication system for preventing  eavesdropping. Inspired by the artificial noise (AN)-based security schemes in traditional wireless communication systems, in this paper, AN is introduced into semantic communication systems for the first time to prevent semantic eavesdropping. However, the introduction of AN also poses challenges for the legitimate receiver in extracting semantic information. Recently, denoising diffusion probabilistic models (DDPM) have demonstrated their powerful capabilities in generating multimedia content. Here, the paired pluggable modules are carefully designed using DDPM. Specifically, the pluggable encryption module generates  AN and adds it to the output of the semantic transmitter, while the pluggable decryption module before semantic receiver uses DDPM to generate the detailed semantic information by removing both AN and the channel noise. In the scenario where the transmitter lacks eavesdropper's knowledge, the artificial Gaussian noise (AGN) is used as AN. We first  model a power allocation optimization problem to determine the power of AGN, in which the objective is to minimize the weighted sum of  data reconstruction error of legal link, the mutual information of illegal link, and the channel input distortion. Then, a deep reinforcement learning  framework using deep deterministic policy gradient is proposed to solve the optimization problem. In the scenario where the transmitter is aware of the eavesdropper's knowledge, we propose an AN generation method based on adversarial residual networks (ARN). Unlike the previous scenario, the mutual information term in the objective function is replaced by the confidence of eavesdropper correctly
  retrieving private information. The adversarial residual network is then trained to minimize the modified objective function. The simulation results show that the  diffusion-enabled pluggable encryption module prevents semantic eavesdropping while the pluggable decryption module  achieves the  high-quality semantic communication.

\end{abstract}
\begin{IEEEkeywords}
Artificial noise, diffusion  models, semantic communication,   wireless security.
\end{IEEEkeywords}

\section{Introduction}
 \IEEEPARstart{S}{emantic} communication is a new communication paradigm that aims to transmit the underlying meaning of source instead of the exact bits \cite{luo2022Se,chen2024personalizing}.
As a post-Shannon communication paradigm, semantic communication has a great potential to achieve the classic Shannon's limit. Recently, both academia and industry have recognized semantic communication as a key enabler for the sixth generation (6G) of wireless networks\cite{qin2022se}. The ultimate vision of  6G, powered  by semantic communication, is to enable the machines to extract the meaning of information at the transmitter side and interpret it at the receiver side. Thus, the semantic communication  technology is expected to support the typical applications of next-generation networks, such as human-machine interaction,   metaverse, healthcare, and intelligent communication \cite{yang2023se}.


Recent advancements in deep learning (DL) have  sparked significant research interest in  DL-enabled semantic communication, which can be broadly categorized into three areas: data recovery,  intelligent task execution, and recovery-and-execution. For the data recovery,  the global semantic features are extracted for transmission, where the common  data types include (but not limited to)  text \cite{xie2021deep}, image\cite{huang2023towards}, and speech \cite{weng2021se}.
In the  intelligent task execution, only the  task-relevant semantic features are extracted for the specific intelligent tasks, such as  question answering \cite{liu2023task}, image classification \cite{zhang2023drl,yang2021semanticcommunicationsaitasks}, and speech  recognition \cite{weng2023deep}.
Beyond designs focused solely on data recovery or task execution, emerging services  such as the virtual reality  have driven  the joint design of data recovery and intelligent task execution \cite{lyu2024se,zhang2023deep}.
These pioneering works \cite{xie2021deep,huang2023towards,weng2021se,liu2023task,zhang2023drl,yang2021semanticcommunicationsaitasks,weng2023deep,lyu2024se,zhang2023deep} have showed  DL-based semantic communication can  offer  higher efficiency and reliability than  bit communication.

However, the performance improvement also makes it easier for eavesdroppers to access private semantic information through attacks such as model inversion and attribute inference \cite{liu2023sem}. For example, using model inversion attack, the eavesdropper can train a surrogate model or steal the broadcast semantic receiver to eavesdrop on the original semantic  information. Regarding as the attribute inference attack, the eavesdropper  uses a local eavesdropping model (e.g., a semantic classifier) to infer private attributes, such as gender and image label\cite{he2024secure}. Thus, although DL-based semantic communication achieves the great performance advantages, it also brings  the leakage risk of private semantic information. To address the eavesdropping problem, the semantic communication community has proposed many innovative works with semantic eavesdropping awareness\cite{kozlov2024securesemanticcommunicationwiretap,chen2024nearlyinformationtheoreticallysecure,march2020adv,erdemir2022privacy,zhang2023wireless,luo2023encrypted,chen2023model}. The general outer and inner bounds on the rate-distortion-equivocation region have been derived to characterize the information-theoretic limits for the secure semantic communication \cite{kozlov2024securesemanticcommunicationwiretap}.  Concurrently, from an information-theoretic security perspective, a superposition code method is proposed to ensure that the eavesdropper decoding is nearly equivalent to random guessing \cite{chen2024nearlyinformationtheoreticallysecure}.  Moreover, given the data-driven nature of the DL-based semantic communication, extensive efforts adopt the adversarial training to  minimize the distortion of legitimate communication link while maximizing that of the eavesdropper's link  \cite{march2020adv,erdemir2022privacy,zhang2023wireless,luo2023encrypted}.
Instead of  adversarial training, the work \cite{chen2023model} suggests using the random permutation and substitution to combat the model inversion attack. Notably, despite the significant efforts made in existing works to counter semantic eavesdropping, several challenges remain that require further attention\cite{liu2023sem,he2024secure,tang2024securesemanticcommunicationimage,du2023rethinking,he2023anti}. These are discussed below.

\begin{itemize}
  \item  {\textbf{Over-distortion of channel input:}} Due to the security design for the semantic communication system, the transmitted signal often suffers from an  over-distortion compared to the original channel input, which may arouse the suspicion of eavesdroppers\cite{he2024secure,tang2024securesemanticcommunicationimage}. In response, eavesdroppers can initiate jamming attacks to disrupt legitimate communications. For example, the use of superposition coding \cite{chen2024nearlyinformationtheoreticallysecure} substantially alters the original modulation scheme, while adversarial training \cite{march2020adv,erdemir2022privacy,zhang2023wireless,luo2023encrypted} and random perturbations \cite{chen2023model} introduce significant changes to the channel input. These easily detectable alterations provide eavesdroppers with new opportunities to launch attacks.
\item {\textbf{Extra computation overhead by retraining:}} When facing eavesdropping threats, it is difficult   to retrain all  semantic communication modules by interrupting the online semantic communication system\cite{liu2023sem,he2024secure}.  For example, in the advanced  adversarial training schemes \cite{march2020adv,erdemir2022privacy,zhang2023wireless,luo2023encrypted},  the semantic communication modules are retrained for  the specific communication and eavesdropping agents.
\item {\textbf{Unknown eavesdropper's prior knowledge:}} The eavesdroppers in semantic communication systems are not always active\cite{he2023anti}. When the passive eavesdropper remains silent and intentionally  hide its presence from the transmitter, it becomes challenging to acquire any prior knowledge about the passive eavesdroppers\cite{nguyen2014secrecy,ng2014robust}. Thus, it is essential to design secure semantic communication systems that account for various practical scenarios, including cases where the transmitter is either aware or unaware of the eavesdropper's knowledge.
\end{itemize}

These challenges underline the practical considerations in deploying security-aware semantic communication systems. Traditional methods often fall short in addressing these layered issues efficiently.  Recently, generative artificial intelligence (GAI) has shown its great potential in creating new contents such as video, pictures, and music.  Leveraging the powerful capabilities of GAI, products like DALL-E $3$ \cite{BetkerImprovingIG}  and ChatGPT \cite{wu2023brief} obtain widespread popularity among consumers. Among these advancements, denoising diffusion probabilistic models (DDPM)  \cite{ho2020denoising}, a class of powerful generative models,  have gained attention for their exceptional ability to generate high-quality data, particularly in image synthesis. The key idea behind DDPM is to model the process of data generation as the reverse of a noise diffusion process. In the forward process, noise is gradually added to data (e.g., images) over a series of time steps until the data becomes indistinguishable from pure noise. This forward process is designed to be simple and tractable, typically implemented as a Gaussian noise injection at each step. Given a data distribution
${\bm{x}}^{(0)}\thicksim q\left({\bm{x}}^{(0)}\right)$, the generated sample at the $t$-th time-step of the forward process can be expressed by
$
{\bm{x}}^{(t)}=\sqrt{\bar{\alpha}^{(t)}}{\bm{x}}^{(0)}+\sqrt{1-\bar{\alpha}^{(t)}}\bm{\epsilon}^{(t)}
$,
where $\bar{\alpha}^{(t)}=\prod_{i=1}^{t}\alpha^{(i)}$, $\alpha^{(i)}\in(0,1)$, and $\bm{\epsilon}^{(t)}$ follows the standard Gaussian distribution.
The reverse process, which is the core of DDPM, aims to progressively denoise the noisy data to generate the original data distribution.
\begin{figure*}[tbp]
  \centering
  \includegraphics[width=4.2in]{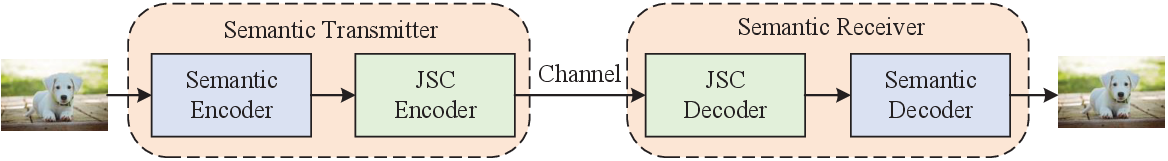}\\
  \caption{ Semantic communication system, where ``JSC" denotes joint source channel. }  \label{SC_system model}
\end{figure*}
In this paper,  we propose a novel diffusion-enabled secure semantic communication against eavesdropping, where the artificial  noise (AN) and the channel noise are mapped to the forward process of diffusion models, and the reverse process of  diffusion models is designed to  adaptively generate  the  detailed semantic information from the received  noisy  signal with  AN  and natural noise (say, channel noise). Given the diffusion-enabled secure semantic communication, we need to carefully design the AN including the noise type and power to satisfy that the diffusion models can generate the semantic information well at Bob while ensuring that the eavesdropper Eve cannot obtain the private information. Specifically, to address the challenge of unknown eavesdropper's prior knowledge, we consider two practical scenarios: one where the transmitter is unaware of the eavesdropper's knowledge and another where the transmitter has the eavesdropper's knowledge.\footnote{Here, eavesdropper's knowledge refers to the DL  model architecture, parameters, and the eavesdropping task.} For the scenario where the transmitter lacks the knowledge of  eavesdropper,  a pluggable  encryption module is designed to generate a low-power  artificial Gaussian noise (AGN), which is added on the output of semantic transmitter with only slight distortion to the channel input. The power of AGN is determined by the deep deterministic policy gradient (DDPG).
Furthermore, we propose to  design a pluggable  decryption module via DDPM to  adaptively generate the semantic information from the received  noisy signal with the  AGN and the channel noise. For the scenario where the transmitter has the knowledge
of  eavesdropper,   the generated AN at the pluggable  encryption
module  is  the low-power
adversarial perturbation, which is obtained using the end-to-end trained adversarial residual
networks (ARN).  Similarly, the  pluggable  decryption module uses DDPM to generate
the semantic information from the received noisy signal with the  adversarial perturbation and  the channel noise.\footnote{The \textit{``pluggable"} design concept  in the security-aware semantic communication system avoids the requirement for retraining the entire system. Specifically, in the presence of eavesdropping risks, such as when the transmitted message of Alice contains private information, the pluggable modules can be plugged; however, in the absence of eavesdropping concerns, as with the public information transmission, these modules can be unplugged.  In other words, the pluggable design enables the online semantic communication system
to dynamically plug and unplug the paired modules to meet customized  security requirements, facilitating the practical deployment in the real world.} Finally, the major contributions  of our work are summarized as follows:


\begin{itemize}
 \item We propose a novel diffusion-enabled secure semantic communication system, which incorporates AN (say,  Gaussian noise or adversarial perturbation) and channel noise into the forward process of diffusion, and uses the adaptive reverse process to  generate
     the semantic information from the received noisy signal with the actively introduced  AN and the  passively introduced natural noise.  Different from the traditional scheme,  we   carefully design the power of AGN in the forward diffusion process   and  explore the positive side of adversarial perturbations, i.e., using ARN to generate adversarial perturbations for preventing  eavesdropping.

 \item We design a security-aware semantic communication framework via the paired pluggable modules including the pluggable encryption module and the paired pluggable decryption module. The pluggable encryption module is designed to generate AN. The paired pluggable decryption module uses  DDPM to generate the semantic information from the received signal with  the AN and the channel noise. Compared to the existing scheme, the designed paired pluggable modules  enable the semantic communication system to operate without the requirement for retraining in response to security threats.

  \item We develop a triple-objective optimization problem for the AGN power allocation  with unknown eavesdropper's knowledge, jointly minimizing the legitimate signal reconstruction error, the eavesdropping link mutual information, and the channel input
distortion. To address the  non-convex problem, we first  derive an upper bound on the mutual information, and then devise a  deep reinforcement learning (DRL) architecture via  DDPG that enables adaptive AGN power generation. The lightweight actor network achieves the fast environment-aware optimization of AGN power.

  \item We propose a novel AN generation method via ARN  with known eavesdropper's knowledge, employing the adversarial training to jointly minimize: 1)  the legitimate signal reconstruction error, 2)  the confidence of eavesdropper correctly retrieving private information, and 3) the channel input distortion. The ARN is optimized using  end-to-end training, which can generate the adversarial examples for any output of the semantic transmitter.

  \item  We  perform extensive experiments on  MNIST, CIFAR-$10$, and Fashion MNIST datasets to evaluate the proposed schemes. Numerical results  show that: 1) the   diffusion-enabled schemes can well balance the communication quality of the legitimate link, the privacy leakage of  illegitimate link, and the  covertness  of  security measures (say, a slight distortion of channel input); 2) the proposed schemes can be applied to scenarios where  eavesdropper's prior knowledge is either known or unknown.
\end{itemize}

The remainder of this paper is organized as follows. Section \ref{SectionII} introduces the general semantic communication system. Security-aware semantic communication framework via the paired pluggable modules is proposed in Section \ref{SectionIII}.
The diffusion-enabled secure pluggable modules are  designed in Section \ref{SectionIV}. Section \ref{SectionV} analyzes the performance of the proposed schemes. Section \ref{SectionVI} concludes this paper.

\begin{figure*}[tbp]
  \centering
  \includegraphics[width=4.6in]{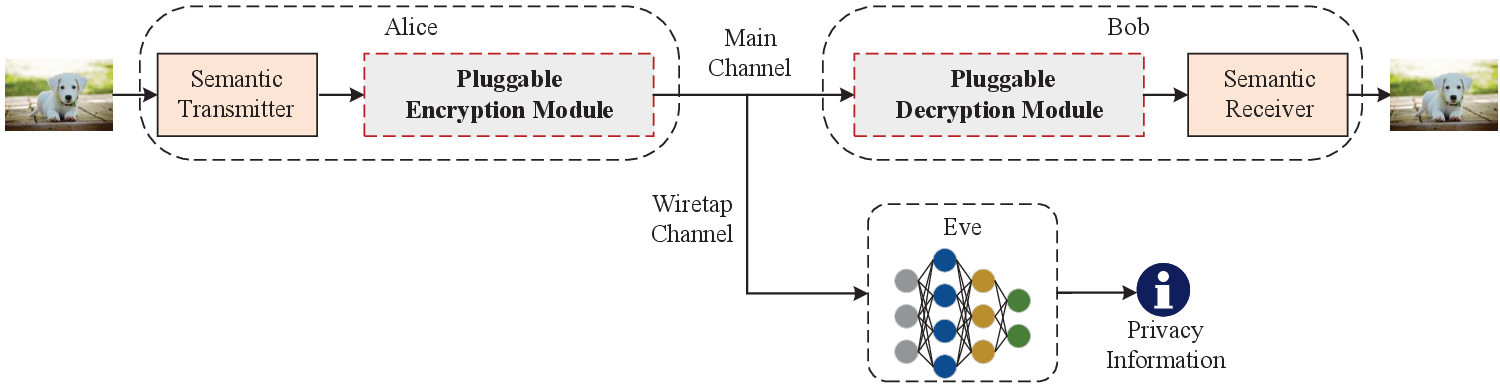}\\
  \caption{ Security-aware semantic communication framework via the paired pluggable modules, where the semantic transmitter consists of semantic encoder and joint source channel encoder, and the semantic receiver includes semantic decoder and joint source channel decoder.  When Alice transmits semantically private information, the paired modules are inserted to prevent eavesdropping; conversely, when transmitting public information, the modules can be unplugged.}  \label{SSC_system model}
\end{figure*}




\section{General Semantic Communication System}\label{SectionII}
Consider a general semantic communication system, as shown in Fig. \ref{SC_system model}, which consists of  semantic encoder (SE), joint source channel   encoder (JSCE),   joint source channel decoder (JSCD), and semantic decoder (SD). The semantic information $\bm{{s}}'$ is extracted as
\begin{align}
\bm{{s}}'=\mathcal{G}_{\text{SE}}(\bm{s}; \theta_{\text{SE}}),
\end{align}
where $\bm{s}\in\mathbb{R}^{L_\text{s}}$ denotes the source data with the length $L_\text{s}$ and $\mathcal{G}_{\text{SE}}(\cdot; \theta_{\text{SE}})$ is the semantic encoder network with the parameter $\theta_{\text{SE}}$. Due to the limited resource, the  semantic information $\bm{{s}}'$ is compressed by
\begin{align}
\bm{x}=\mathcal{G}_{\text{JSCE}}(\bm{{s}}'; \theta_{\text{JSCE}}),
\end{align}
where $\bm{x}\in\mathbb{R}^{L_\text{t}}$ is the transmitted signal with the length $L_\text{t}<L_\text{s}$ and $\mathcal{G}_{\text{JSCE}}(\cdot; \theta_{\text{JSCE}})$ is the joint source channel encoder network with the parameter $\theta_{\text{JSCE}}$. Here, the joint source channel encoder is used to compress the semantic
information for signal transmission and improve the robustness to noise. The received signal $\bm{y}$ is expressed by
\begin{align}
\bm{y}=\bm{x}+\bm{n},
\end{align}
where $\bm{n}$ is the white noise, which follows the zero-mean Gaussian distribution with the variance $\sigma^2\bm{I}_{L_{\text{t}}}$. Using the joint source channel  decoder, the semantic information can be recovered as
\begin{align}
\hat{{\bm{s}}}'=\mathcal{D}_{\text{JSCD}}(\bm{y};\theta_{\text{JSCD}}),
\end{align}
where $\hat{{\bm{s}}}'$ is the recovered semantic feature  and $\mathcal{D}_{\text{JSCD}}(\cdot;\theta_{\text{JSCD}})$ is the joint source channel decoder with the parameter $\theta_{\text{JSCD}}$. The task of  joint source channel decoder is to decompress the semantic information while mitigating the effects of channel distortion. Finally, the recovered semantic representation is  fed into the semantic decoder to obtain the original data,  which can be given by
\begin{align}
\hat{\bm{s}}=\mathcal{D}_{\text{SD}}(\hat{{\bm{s}}}';\theta_{\text{SD}}).
\end{align}

Here, we take the image transmission task as an example.  To jointly train the encoder and the decoder, we define the  mean square of error (MSE) as the loss function, i.e.
\begin{align}\label{SC_MSE}
\mathcal{L}({\varTheta})\triangleq \mathbb{E}\|\bm{s}- \hat{\bm{s}}\|^2,
\end{align}
where $\varTheta\triangleq\{\theta_{\text{SE}}, \theta_{\text{JSCE}}, \theta_{\text{JSCD}}, \theta_{\text{SD}}\}$ denotes  the set of all trainable
parameters of the semantic communication system.  Mathematically, the set of all trainable parameters can be obtained by solving the following optimization problem, i.e.
\begin{align} \label{SCopti}
\mathop{\text{min}}_{\varTheta}\:\mathcal{L}({\varTheta}),
\end{align}
where the problem \eqref{SCopti} can be solved using the  stochastic gradient descent algorithm. Once the parameter set $\varTheta$ is obtained, the semantic transmitter and the semantic receiver are deployed in the practical image transmission application.

\begin{remark}
We want to emphasize that, for convenience, we model the semantic communication system specifically for image transmission and use MSE as the loss function. However, the  formulated system can be extended to other tasks with only simple modifications to the loss function.
\end{remark}

\section{Security-aware Semantic Communication Framework via Paired Pluggable Modules} \label{SectionIII}
In this section, we will discuss the eavesdropping threats faced in semantic communication system of Section \ref{SectionII}. Following  the naming convention in the security research,
Alice, Bob, and Eve are used to represent  the sender, the legitimate receiver, and
the eavesdropper, respectively. As shown in Fig. \ref{SSC_system model},  Alice wants to reliably transmit source data to Bob, while Eve tries to eavesdrop on the privacy information of Alice. To degrade the information flow to Eve, Alice adds an additional pluggable  encryption module after the semantic transmitter. Then, the transmitted signal of Alice can be denoted by
\begin{align}
\bm{x}_{\text{Alice}}'=f_{\text{Alice}}({\bm{x}}_{\text{Alice}}),
\end{align}
where $f_{\text{Alice}}(\cdot)$ is the   pluggable  encryption module to be designed and ${\bm{x}}_{\text{Alice}}$ is the output of the semantic transmitter, which is expressed by
 \begin{align}
 {\bm{x}}_{\text{Alice}}=\mathcal{G}_{\text{JSCE}}\left(\mathcal{G}_{\text{SE}}(\bm{s}_{\text{Alice}})\right),
 \end{align}
where $\bm{s}_{\text{Alice}}$ is source data of Alice. For convenience, ${\bm{x}}_{\text{Alice}}$ is normalized to a power of $1$.
Accordingly, the received signal $\bm{y}_{\text{Bob}}$ of Bob  is represented as
\begin{align}
\bm{y}_{\text{Bob}}=\bm{x}_{\text{Alice}}'+\bm{n}_{\text{Bob}},
\end{align}
and the received signal $\bm{y}_{\text{Eve}}$ of Eve   can be written as
\begin{align}
\bm{y}_{\text{Eve}}=\bm{x}_{\text{Alice}}'+\bm{n}_{\text{Eve}},
\end{align}
where $\bm{n}_{\text{Bob}}$ and $\bm{n}_{\text{Eve}}$ are the white noises, which  follow the zero-mean Gaussian distribution with the variances $\sigma^2_{\text{Bob}}\bm{I}_{L_{\text{t}}}$ and $\sigma^2_{\text{Eve}}\bm{I}_{L_{\text{t}}}$, respectively.\footnote{In this paper, we  focus on the AWGN channel to enhance the reader's understanding of the approaches proposed in the subsequent sections. Notably, by modeling the physical channel as a neural network \cite{xie2021deep}, our findings can be extended to more complex channel models. Furthermore, when both legitimate and illegitimate links are characterized as AWGN channels, designing an effective security scheme becomes particularly challenging, as the differences between the channels can no longer be leveraged to mitigate  eavesdropping.}  Because Alice-Bob link is the legitimate communication link, Bob uses a paired  pluggable decryption module in front of the semantic receiver. The output $\bm{x}_{\text{Bob}}$ of the pluggable   decryption module is given by
\begin{align}
\bm{x}_{\text{Bob}}=f_{\text{Bob}}(\bm{y}_{\text{Bob}}),
\end{align}
where $f_{\text{Bob}}(\cdot)$ is the pluggable  decryption module
to be designed. Finally, the recovered  data $\bm{s}_{\text{Bob}}$  is expressed as
\begin{align}
\bm{s}_{\text{Bob}}=\mathcal{D}_{\text{SD}}\left(\mathcal{D}_{\text{JSCD}}(\bm{x}_{\text{Bob}})\right).
\end{align}
Meanwhile, the eavesdropper Eve attempts to infer the private information from $\bm{y}_{\text{Eve}}$ using its own neural network $g_{\text{Eve}}(\cdot;\theta_{\text{Eve}})$ with the parameter $\theta_{\text{Eve}}$. The obtained  private information at Eve can be denoted by
\begin{align}
\bm{s}_\text{Eve}=g_{\text{Eve}}(\bm{y}_{\text{Eve}};\theta_{\text{Eve}}).
\end{align}

From Alice's  perspective, if Eve is a passive eavesdropper, remaining silent throughout, Alice cannot acquire the knowledge of Eve. Conversely, if Eve is an active eavesdropper, impersonating a legitimate user participating in the entire wireless communication network, Alice can obtain the  knowledge of Eve. In this paper, we consider two  practical scenarios:
\begin{itemize}
 \item \textbf{Scenario I:} Alice lacks the knowledge of Eve's neural network $g_{\text{Eve}}(\cdot;\theta_{\text{Eve}})$.
     \item \textbf{Scenario II:} Alice has the knowledge of Eve's neural network $g_{\text{Eve}}(\cdot;\theta_{\text{Eve}})$.
 \end{itemize}

The goal of this paper is to design the paired pluggable   modules (say, $f_{\text{Alice}}$ and $f_{\text{Bob}}$) without training the deployed semantic transceiver, so as to ensure the secure  transmission when facing eavesdropping  threat. In Scenario I, since Alice has no the prior knowledge about Eve, we design a pluggable  encryption module that generates the  AGN. Correspondingly, a paired pluggable decryption module is designed to generate
the semantic information from the received noisy signal with  the actively introduced AGN and the channel noise. In Scenario II, because Alice possesses the prior knowledge about Eve, we design a pluggable encryption module that generates the adversarial perturbation tailored specifically to Eve. Similarly, a paired pluggable decryption module is designed to  generate
the semantic information from the received noisy signal with the actively introduced adversarial perturbation and the channel noise. The specific designs for Scenarios I and II will be shown in Sections \ref{no_eve_know} and \ref{has_eve_know} in detail, respectively.

\begin{remark}
In the designed secure semantic communication system, the paired pluggable modules are plugged in and out according to the security requirements.\footnote{Note that, in our work, the pluggable modules  are designed  from the perspective of semantic communication security, with the paired modules tailored to specific tasks. In our future work, we will also explore task-agnostic  designs for the pluggable modules.} Specifically, when Alice transmits the signal that contains private information, these modules are plugged; conversely, they are unplugged.
\end{remark}

\section{Diffusion-enabled Secure Paired Pluggable Modules Design} \label{SectionIV}
In this section, we first formulate a general pluggable modules optimization problem for the security-aware semantic communication system. Then, the  pluggable modules via diffusion models are carefully designed for the practical scenarios where Alice either possesses or lacks the knowledge of Eve. The designed  modules  serve as a crucial components
for the security-aware semantic communication system by adding and removing AN.

\subsection{Problem Formulation}
In the security-aware semantic communication system, the ultimate goal is to simultaneously ensure the semantic communication quality of the Alice-Bob link, protect the semantic information transmitted by Alice from being disclosed to Eve, and avoid the over-distortion of channel input (i.e., covertness). Mathematically, the general optimization problem for the security-aware semantic communication system can be  formulated by
\begin{align} \label{SSC_problem}
&\mathop{\text{min}}_{f_{\text{Alice}}, f_{\text{Bob}}}\: \lambda_{\text{com}} d_{\text{com}}(\bm{s}_{\text{Alice}}, \bm{s}_{\text{Bob}})+\lambda_{\text{pri}} d_{\text{pri}}(\bm{s}_{\text{Alice}}, \bm{s}_{\text{Eve}})\notag\\
&~~~~~~~~~~~~~~~~~~~~~~~~~~~+\lambda_{\text{cov}} d_{\text{cov}}(\bm{x}_{\text{Alice}}, \bm{x}_{\text{Alice}}'),
\end{align}
where $d_{\text{com}}(\cdot)$ is the task-specific function that characterizes the communication quality of the Alice-Bob link, depending on the specific task of semantic communication; $d_{\text{pri}}(\cdot)$ describes the privacy leakage of the Alice-Eve link; $d_{\text{cov}}(\cdot)$ characterizes the covertness of the security measures;\footnote{If the difference in channel input is significant before and after Alice employs  encryption module $f_{\text{Alice}}$, Eve can easily detect that Alice has implemented some security measures. For example, $\bm{x}_{\text{Alice}}'$ has a significant power change relative to $\bm{x}_{\text{Alice}}$. In this case, Eve may potentially engage in more sophisticated eavesdropping or launch active attacks on Alice-Bob link.}
$\lambda_{\text{com}}$, $d_{\text{pri}}$, and $d_{\text{cov}}$ are the hyper-parameters that control the balance of the multiple objectives.   In the following  Sections \ref{no_eve_know} and \ref{has_eve_know}, we will show how to design $f_{\text{Alice}}$ and $f_{\text{Bob}}$ in detail.

\subsection{Secure Paired Pluggable Modules Design for Alice having no Eve's knowledge} \label{no_eve_know}
As mentioned in Section \ref{SectionII}, our primary focus for Alice-Bob link is on the data reconstruction task.\footnote{The proposed scheme  can be directly extended to other tasks. Here, we use the data reconstruction task as a typical example to illustrate our approach.} Thus, the metric function $d_{\text{com}}(\bm{s}_{\text{Alice}}, \bm{s}_{\text{Bob}})$ can be defined as
\begin{align}
d_{\text{com}}(\bm{s}_{\text{Alice}}, \bm{s}_{\text{Bob}})&\triangleq\mathbb{E}\|\bm{s}_{\text{Alice}}-\bm{s}_{\text{Bob}}\|^2, \label{communication}
\end{align}
where the equation \eqref{communication} is \emph{communication MSE} between the source data $\bm{s}_{\text{Alice}}$ and the recovery data $\bm{s}_{\text{Bob}}$, in which a smaller value indicates a better data reconstruction. It is worth noting that MSE is not mandatory and depends on the specific task. For example, in image classification tasks, metrics such as classification accuracy can be used as measures to evaluate the quality of legitimate communication. To quantify the privacy leakage of the Alice-Eve link, we define $d_{\text{pri}}(\bm{s}_{\text{Alice}}, \bm{s}_{\text{Eve}})$ as the  MI function, which does not require knowledge of neural network $g_{\text{Eve}}(\cdot;\theta_{\text{Eve}})$. Specifically,  the metric function  $d_{\text{pri}}(\bm{s}_{\text{Alice}}, \bm{s}_{\text{Eve}})$ is given by
\begin{align}
d_{\text{pri}}(\bm{s}_{\text{Alice}}, \bm{s}_{\text{Eve}})&\triangleq I(\bm{x}_{\text{Alice}};\bm{y}_{\text{Eve}}),\label{privacy}
\end{align}
where the equation \eqref{privacy} denotes the \emph{privacy leakage MI} between  $\bm{x}_{\text{Alice}}$ and $\bm{y}_{\text{Eve}}$, in which a smaller value indicates less information leaked to Eve\cite{kozlov2024securesemanticcommunicationwiretap,erdemir2022privacy}. Here, the privacy leakage MI  denotes the achievable rate of the Alice-Eve link, which reflects the loss of
 privacy information due to the pluggable encryption module  $f_{\text{Alice}}(\cdot)$  and the channel noise. For the covertness of the security measures, the metric function $d_{\text{cov}}(\bm{x}_{\text{Alice}}, \bm{x}_{\text{Alice}}')$ is defined as
\begin{align}
d_{\text{cov}}(\bm{x}_{\text{Alice}}, \bm{x}_{\text{Alice}}')&\triangleq\mathbb{E}\|\bm{x}_{\text{Alice}}-\bm{x}_{\text{Alice}}'\|^2,\label{perceptibility}
\end{align}
where the equation \eqref{perceptibility} is \emph{covertness MSE} between the input $\bm{x}_{\text{Alice}}$   and the output $\bm{x}_{\text{Alice}}'$ of the pluggable encryption module, in which  a smaller value suggests less channel input distortion and thus better  covertness for $f_{\text{Alice}}$. From the above specific metrics, the problem \eqref{SSC_problem} can be transformed to
\begin{align} \label{SSC_problem_trans}
&\mathop{\text{min}}_{f_{\text{Alice}}, f_{\text{Bob}}}\:\lambda_{\text{com}}\mathbb{E}\|\bm{s}_{\text{Alice}}-\bm{s}_{\text{Bob}}\|^2+\lambda_{\text{pri}} I(\bm{x}_{\text{Alice}};\bm{y}_{\text{Eve}})\notag\\
&~~~~~~~~~~~~~~~~~~~~~+\lambda_{\text{cov}} \mathbb{E}\|\bm{x}_{\text{Alice}}-\bm{x}_{\text{Alice}}'\|^2.
\end{align}

Next,  our goal is to solve  the problem \eqref{SSC_problem_trans}, where the core challenge is to carefully design $f_{\text{Alice}}$ and $f_{\text{Bob}}$.

\subsubsection{ Design of $f_{\text{Alice}}$} Inspired by the classic security schemes using artificial noise, we first define  the pluggable encryption module $f_{\text{Alice}}$ as
\begin{align}
\bm{x}_{\text{Alice}}'&=f_{\text{Alice}}({\bm{x}}_{\text{Alice}};u)\\
&=\sqrt{u}{\bm{x}}_{\text{Alice}}+\sqrt{1-u}\bm{w}, \label{AN_alice}
\end{align}
where $u$ is the power allocation factor to be designed and $\bm{w}$ is the AGN satisfying $\bm{w} \thicksim\mathcal{N}(\bm{0},\bm{I}_{L_{\text{t}}})$.

\subsubsection{ Design of $f_{\text{Bob}}$} From \eqref{AN_alice}, the received signals of Bob and Eve can be respectively expressed by
\begin{align}
\bm{y}_{\text{Bob}}&=\sqrt{u}{\bm{x}}_{\text{Alice}}+\sqrt{1-u}\bm{w}+\bm{n}_{\text{Bob}}, \label{diff_Bob}\\
\bm{y}_{\text{Eve}}&=\sqrt{u}{\bm{x}}_{\text{Alice}}+\sqrt{1-u}\bm{w}+\bm{n}_{\text{Eve}}.\label{diff_Eve}
\end{align}
Evidently, the total amount of information leaked to Eve vanish as $u\rightarrow0$, i.e.
\begin{align} \label{mul_sz}
I(\bm{x}_{\text{Alice}};\bm{y}_{\text{Eve}})\mathop{\longrightarrow}\limits_{u\rightarrow0} 0.
\end{align}
However, $u\rightarrow0$ also deteriorates the quality of  Alice-Bob link, which presents a challenge for Bob to reliably restore the data. To address the challenge,  the design principle of $f_{\text{Bob}}$ is to generate the semantic information from the received noisy signal with the actively introduced AGN  and the   passively introduced channel noise. Inspired by DDPM, we map the signal transmission process to the \emph{forward diffusion process}, and the denoising process to the \emph{inverse diffusion process}.

{\bf{AGN and channel noise as the part of  forward diffusion process:}}  Let ${\bm{x}}_{\text{Alice}}^{(0)}\thicksim q\left({\bm{x}}_{\text{Alice}}^{(0)}\right)$ be the original signal ${\bm{x}}_{\text{Alice}}$ generated by the semantic transmitter, where $q\left({\bm{x}}_{\text{Alice}}^{(0)}\right)$ denotes the distribution of ${\bm{x}}_{\text{Alice}}^{(0)}$. The forward diffusion process is a Markov process, which is defined by gradually adding the Gaussian noise at each time-step $t$, i.e.
\begin{align}\label{FDPR}
{\bm{x}}_{\text{Alice}}^{(t)}=\sqrt{1-\beta^{(t)}}{\bm{x}}_{\text{Alice}}^{(t-1)}+\sqrt{\beta^{(t)}}\bm{\epsilon}^{(t-1)},
\end{align}
where $\bm{\epsilon}^{(t-1)}\thicksim\mathcal{N}(\bm{0}, \bm{I})$,  $0<\beta^{(t)}<1$ is the known variance, and $t$ denotes the time index. Thus, given a maximum time-step $T$, the joint probability distribution of the forward process is written as
\begin{align} \label{FDP}
q\left({\bm{x}}_{\text{Alice}}^{(1)},\dots, {\bm{x}}_{\text{Alice}}^{(T)} |{\bm{x}}_{\text{Alice}}^{(0)}\right)=\Pi_{t=1}^{T}q\left({\bm{x}}_{\text{Alice}}^{(t)}|{\bm{x}}_{\text{Alice}}^{(t-1)}\right),
\end{align}
where
\begin{align} \label{FDPsub}
q\left({\bm{x}}_{\text{Alice}}^{(t)}|{\bm{x}}_{\text{Alice}}^{(t-1)}\right)
\thicksim
\mathcal{N}\left({\bm{x}}_{\text{Alice}}^{(t)};\sqrt{1-\beta^{(t)}}{\bm{x}}_{\text{Alice}}^{(t-1)},\beta^{(t)}\bm{I}\right).
\end{align}
Resorting to the reparameterization trick, the equation \eqref{FDPR} can be transformed to
\begin{align} \label{repara}
{\bm{x}}_{\text{Alice}}^{(t)}=\sqrt{\bar{\alpha}^{(t)}}{\bm{x}}_{\text{Alice}}^{(0)}+\sqrt{1-\bar{\alpha}^{(t)}}\bm{\epsilon}^{(t)},
\end{align}
where $\bar{\alpha}^{(t)}=\prod_{i=1}^{t}\alpha^{(i)}$ and $\alpha^{(t)}=1-\beta^{(t)}$. Comparing \eqref{diff_Bob} and \eqref{repara}, we can map  AGN and  channel noise to the part of the forward diffusion process.

{\bf{Adaptive inverse diffusion  process:}}  Using the Bayes rule, we can derive $q\left({\bm{x}}_{\text{Alice}}^{(t-1)}|{\bm{x}}_{\text{Alice}}^{(t)}\right)$ as
\begin{align}
q\left({\bm{x}}_{\text{Alice}}^{(t-1)}|{\bm{x}}_{\text{Alice}}^{(t)}\right)
\thicksim \mathcal{N}\left({\bm{x}}_{\text{Alice}}^{(t-1)};\tilde{\bm{\mu}}\left({\bm{x}}_{\text{Alice}}^{(t)},t\right),\tilde{\beta}^{(t)}\bm{I}\right),
\end{align}
where
\begin{align}
\tilde{\bm{\mu}}\left({\bm{x}}_{\text{Alice}}^{(t)},{\bm{x}}_{\text{Alice}}^{(0)},t\right)&=
\frac{1}{\sqrt{\alpha^{(t)}}}\left({\bm{x}}_{\text{Alice}}^{(t)}-\frac{1-\alpha^{(t)}}{\sqrt{1-\bar{\alpha}^{(t)}}}\bm{\epsilon}^{(t)}\right), \label{mu_bar}\\
\tilde{\beta}^{(t)}&=\frac{1-\bar{\alpha}^{(t-1)}}{1-\bar{\alpha}^{(t)}}\beta^{(t)}. \label{beta_bar}
\end{align}
The  noise estimation neural  network (NENN) $f_{\text{NENN}}(\cdot; \theta_{\text{NENN}})$ with the parameter $\theta_{\text{NENN}}$ is used to predict the noise $\bm{\epsilon}^{(t)}$, where the loss function of training phase is defined as
\begin{align} \label{loss_NENN}
&\mathcal{L}_{\text{NENN}}(\theta_{\text{NENN}})\notag\\
&=\mathbb{E}_{t,{\bm{x}}_{\text{Alice}}^{(0)},\bm{\epsilon}^{(t)}}
\left\|\bm{\epsilon}^{(t)}-f_{\text{NENN}}\left({\bm{x}}_{\text{Alice}}^{(t)},t;\theta_{\text{NENN}}\right)\right\|^2,
\end{align}
where ${\bm{x}}_{\text{Alice}}^{(t)}$ is generated using \eqref{repara} and
$t$ is sampled uniformly from $\{1, 2, \dots, T\}$.
\begin{figure*}[tbp]
  \centering
  \includegraphics[width=4.7in]{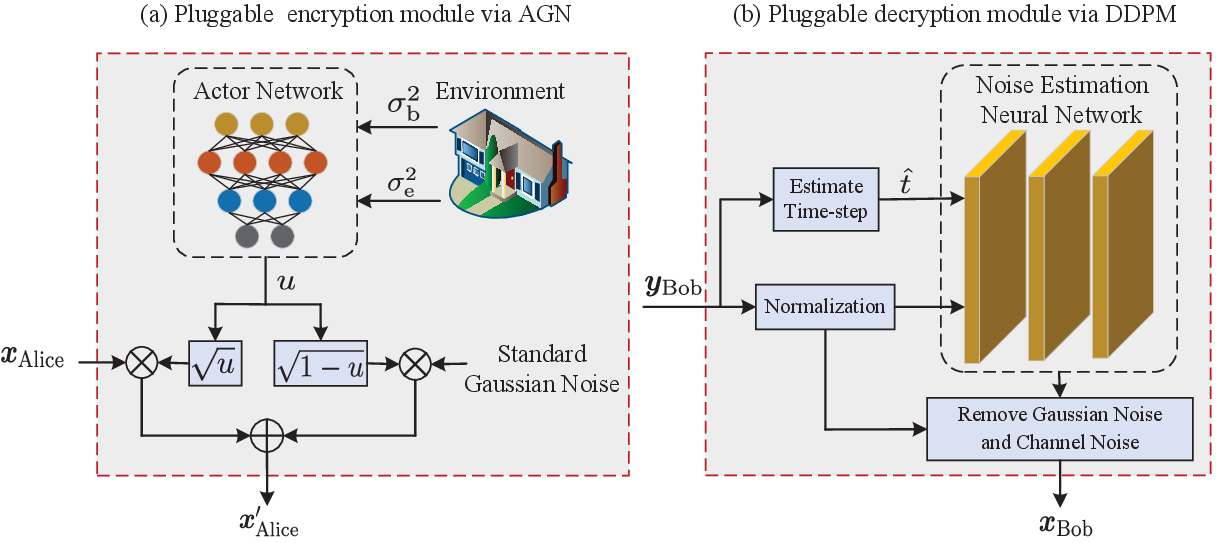}\\
  \caption{  Pluggable paired modules for Alice having no Eve's knowledge, where (a) is the pluggable encryption module via AGN, and (b) is the pluggable  decryption module via DDPM.}  \label{AN}
\end{figure*}

Once the optimal parameter $\theta_{\text{NENN}}^{*}$ of  NENN is obtained,  the pluggable decryption module $f_{\text{Bob}}$ can be designed as
\begin{align}
{\bm{x}}_{\text{Bob}}&= f_{\text{Bob}}({\bm{y}_{\text{Bob}}};\theta_{\text{NENN}}^{*})\\
&=\frac{1}{\sqrt{\bar{\alpha}^{(\hat{t})}}}\left(\bar{\bm{y}}_{\text{Bob}}
-\sqrt{1-\bar{\alpha}^{(\hat{t})}}f_{\text{NENN}}(\bar{\bm{y}}_{\text{Bob}},\hat{t};\theta_{\text{NENN}}^{*})\right), \label{esti_x_bob}
\end{align}
where the equation \eqref{esti_x_bob} follows from  \eqref{repara}; $\bar{\bm{y}}_{\text{Bob}}$ denotes the  normalized version of ${\bm{y}}_{\text{Bob}}$, i.e., $\bar{\bm{y}}_{\text{Bob}}=\frac{{\bm{y}}_{\text{Bob}}}{\sqrt{1+\sigma^{2}_\text{b}}}$;
 the estimated time-step is obtained by
\begin{align}
\hat{t}=\mathop{\text{argmin}}_{t}\left\|\frac{u}{1-u+\sigma^{2}_\text{b}}-\frac{\bar{\alpha}^{(t)}}{1-\bar{\alpha}^{(t)}}\right\|^2, \label{T_tilde}
\end{align}
where the equation \eqref{T_tilde} aims  to find a time-step $t$ such that the signal-to-noise ratio (SNR) of forward diffusion process closely matches that of the Alice-Bob link. As a result,  the amount of noise introduced during the forward diffusion process in equation \eqref{repara}  is kept as consistent as possible with the noise introduced during the signal transmission process in equation \eqref{diff_Bob}.



\subsubsection{Power allocation} So far, we have designed  $f_{\text{Alice}}$ and $f_{\text{Bob}}$ for the case of Alice having no Eve's knowledge, as shown in \eqref{AN_alice} and \eqref{esti_x_bob}.
Next, our goal is to find an optimal power allocation $u$, i.e.
\begin{align} \label{SSC_problem_trans_u}
&\mathop{\text{min}}_{u}\: \lambda_{\text{com}}\mathbb{E}\|\bm{s}_{\text{Alice}}-\bm{s}_{\text{Bob}}\|^2+\lambda_{\text{pri}} I(\bm{x}_{\text{Alice}};\bm{y}_{\text{Eve}})\notag \\
&~~~~~~~~~~~~~~~~+\lambda_{\text{cov}} \mathbb{E}\|\bm{x}_{\text{Alice}}-\bm{x}_{\text{Alice}}'\|^2.
\end{align}
Solving  problem \eqref{SSC_problem_trans_u} is intractable due to minimizing MI. To address the problem, we first give the following Theorem \ref{theorem1}.

\begin{theorem} \label{theorem1}
 The upper bound of  MI $I(\bm{x}_{\text{Alice}};\bm{y}_{\text{Eve}})$ can be given by
\begin{align} \label{UP}
I_{\text{UP}}(\bm{x}_{\text{Alice}};\bm{y}_{\text{Eve}})&=\mathbb{E}_{p(\bm{x}_{\text{Alice}},\bm{y}_{\text{Eve}})}\log\:p(\bm{y}_{\text{Eve}}|\bm{x}_{\text{Alice}})
\notag\\
&~~-\mathbb{E}_{p(\bm{x}_{\text{Alice}})}\mathbb{E}_{p(\bm{y}_{\text{Eve}})}\log\:p(\bm{y}_{\text{Eve}}|\bm{x}_{\text{Alice}}).
\end{align}
\end{theorem}

\emph{Proof:} The proof idea is to demonstrate that the difference between $I_{\text{UP}}(\bm{x}_{\text{Alice}};\bm{y}_{\text{Eve}})$ and $I(\bm{x}_{\text{Alice}};\bm{y}_{\text{Eve}})$
 is non-negative. The difference is given by
\begin{align}
&I_{\text{UP}}(\bm{x}_{\text{Alice}};\bm{y}_{\text{Eve}})-I(\bm{x}_{\text{Alice}};\bm{y}_{\text{Eve}})\notag\\
&=\mathbb{E}_{p(\bm{x}_{\text{Alice}},\bm{y}_{\text{Eve}})}\log\:p(\bm{y}_{\text{Eve}}|\bm{x}_{\text{Alice}})\notag \\
&-\mathbb{E}_{p(\bm{x}_{\text{Alice}})}\mathbb{E}_{p(\bm{y}_{\text{Eve}})}\log\:p(\bm{y}_{\text{Eve}}|\bm{x}_{\text{Alice}})\notag\\
&-\mathbb{E}_{p(\bm{x}_{\text{Alice}},\bm{y}_{\text{Eve}})}
\left(\log\:p(\bm{y}_{\text{Eve}}|\bm{x}_{\text{Alice}})-\log\:p(\bm{y}_{\text{Eve}})\right)\notag\\
&=\mathbb{E}_{p(\bm{y}_{\text{Eve}})}\left(\log\:p(\bm{y}_{\text{Eve}})-\mathbb{E}_{p(\bm{x}_{\text{Alice}})}\log\:p(\bm{y}_{\text{Eve}}|\bm{x}_{\text{Alice}})\right)\notag\\
&\geq0, \label{proof_up}
\end{align}
where \eqref{proof_up} follows from
\begin{align}
&\log\:p(\bm{y}_{\text{Eve}})-\mathbb{E}_{p(\bm{x}_{\text{Alice}})}\log\:p(\bm{y}_{\text{Eve}}|\bm{x}_{\text{Alice}})\notag\\
&=\log\:\left(\mathbb{E}_{p(\bm{x}_{\text{Alice}})}p(\bm{y}_{\text{Eve}}|\bm{x}_{\text{Alice}})\right)-\mathbb{E}_{p(\bm{x}_{\text{Alice}})}\log\:p(\bm{y}_{\text{Eve}}|\bm{x}_{\text{Alice}})\notag\\
&\geq0, \label{jen}
\end{align}
where \eqref{jen} follows from Jensen's inequality. Thus, we conclude that the gap between the upper bound and the true MI is always non-negative.$\hfill\qedsymbol$


From Theorem \ref{theorem1}, we propose to minimize the upper bound of MI, and then the problem \eqref{SSC_problem_trans_u} is further transformed to
\begin{align} \label{SSC_problem_trans_u_MI}
&\mathop{\text{min}}_{u}\:\lambda_\text{com} \mathbb{E}\|\bm{s}_{\text{Alice}}-\bm{s}_{\text{Bob}}\|^2+\lambda_\text{pri} I_{\text{UP}}(\bm{x}_{\text{Alice}};\bm{y}_{\text{Eve}})\notag\\
&~~~~~~~~~~+\lambda_\text{cov} \mathbb{E}\|\bm{x}_{\text{Alice}}-\bm{x}_{\text{Alice}}'\|^2,
\end{align}
where the logarithm of the conditional probability in $I_{\text{UP}}(\bm{x}_{\text{Alice}};\bm{y}_{\text{Eve}})$ is derived as follows:
\begin{align}
&\log\:p(\bm{y}_{\text{Eve}}|\bm{x}_{\text{Alice}})\notag \\
&=-\frac{L_{\text{t}}}{2}\log{2\pi(\sigma_e^2+1-u)}
-\frac{\|\bm{y}_{\text{Eve}}
-\sqrt{u}\bm{x}_{\text{Alice}}\|^2}{2(\sigma_e^2+1-u)}.
\end{align}

One feasible  way to solve the problem \eqref{SSC_problem_trans_u_MI} is the exhaustive search method. However,
the exhaustive search method suffers from high complexity because it requires multiple evaluations of the objective value.  To address this challenge, we propose  a DRL framework using  DDPG algorithm to solve the problem \eqref{SSC_problem_trans_u_MI}. The primary motivation for this approach lies in the DDPG's capability to efficiently approximate the continuous action spaces, allowing for the rapid determination of the optimal parameter
$u$. This methodology not only enhances the computational efficiency but also adapts to the dynamic  SNR conditions, making it a robust solution for the real-time applications in the complex environments.

 Here, we first give the state space, the action space, and the reward of  the DRL framework for solving the problem \eqref{SSC_problem_trans_u_MI}. Specifically, the state space is defined as $\mathcal{S}\triangleq\{\bm{\varpi}_i\}$, where $\bm{\varpi}_i=[(\sigma_{\text{b}}^2)_i, (\sigma_{\text{e}}^2)_i]^{\mathsf{T}}$ and $i$ denotes the state index. The power allocation factor $u$ is selected as the action, and the action space is defined as $\mathcal{A}\triangleq\{{u}_i\}$. According to our optimization problem \eqref{SSC_problem_trans_u_MI}, the reward is designed by
$\mathcal{R}=\{r_i\}$, which is formulated as
\begin{align} \label{reward}
r_i&=-\lambda_\text{com} \mathbb{E}\|\bm{s}_{\text{Alice}}-(\bm{s}_{\text{Bob}})_i\|^2-\lambda_\text{pri} I_{\text{UP}}(\bm{x}_{\text{Alice}};(\bm{y}_{\text{Eve}})_i)\notag\\
&~~~~~~~~~~-\lambda_\text{cov} \mathbb{E}\|\bm{x}_{\text{Alice}}-(\bm{x}_{\text{Alice}}')_i\|^2.
\end{align}
The DRL framework via DDPG consists of online actor network $\pi(\cdot;\theta_{\pi})$, target actor network $\tilde{\pi}(\cdot;\theta_{\tilde{\pi}})$, online critic network $Q(\cdot;\theta_{Q})$, and target critic network $\tilde{Q}(\cdot;\theta_{\tilde{Q}})$, where $\theta_{\pi}$, $\theta_{\tilde{\pi}}$, $\theta_{Q}$, and $\theta_{\tilde{Q}}$ denote the corresponding network parameters.  The online critic network is updated by solving the following problem:
\begin{align}  \label{update_ocn}
\mathop{\text{min}}_{\theta_Q}\:\mathbb{E}\left(\zeta_i-Q(\bm{\varpi}_i,u_i;\theta_Q)\right)^2,
\end{align}
where $\zeta_i$ is the target value, which is defined as
\begin{align}
\zeta_i=r_i+\gamma\tilde{Q}\left(\bm{\varpi}_{(i+1)},\tilde{\pi}(\bm{\varpi}_{(i+1)};\theta_{\tilde{\pi}});\theta_{\tilde{Q}}\right),
\end{align}
where $\gamma$ is the discount factor. For the online actor network, the objective is to optimize $\theta_{\pi}$ that maximizes the Q-value. The optimization problem for the online actor network can be formulated as
\begin{align} \label{update_oan}
\mathop{\text{max}}_{\theta_{\pi}}\:\mathbb{E}\left(Q(\bm{\varpi}_i,\pi(\bm{\varpi}_i;\theta_{\pi});\theta_Q)\right).
\end{align}
Because the scale factor $u$ is continuously differentiable, we can get the gradient expression about $\theta_{\pi}$ by the following chain rule, i.e.
\begin{align}
&\nabla_{\theta_{\pi}}\mathbb{E}\left(Q(\bm{\varpi}_i,\pi(\bm{\varpi}_i;\theta_{\pi});\theta_Q)\right) \\
&\approx
\mathbb{E}\left(\nabla_{\theta_{\pi}}Q(\bm{\varpi}_i,\pi(\bm{\varpi}_i;\theta_{\pi});\theta_Q)\right)\\
&=\mathbb{E}\left(\nabla_{\pi(\bm{\varpi}_i;\theta_{\pi})}Q(\bm{\varpi}_i,\pi(\bm{\varpi}_i;\theta_{\pi});\theta_Q)\nabla_{\theta_{\pi}}{\pi(\bm{\varpi}_i;\theta_{\pi})}\right).
\end{align}

To improve the training stability,  three  strategies are used in the  training of  DDPG. First,  the target networks have the same structures as the online networks, and the parameters of the target networks are updated by
\begin{align}
\theta_{\tilde{Q}}&\leftarrow \xi\theta_{{Q}}+(1-\xi)\theta_{\tilde{Q}}, \label{tar_Q}\\
\theta_{\tilde{\pi}}&\leftarrow \xi\theta_{{\pi}}+(1-\xi)\theta_{\tilde{\pi}},\label{tar_pi}
\end{align}
where $\xi$ is the soft update coefficient, which is a  constant. Second, the experience set $\{\bm{\varpi}_i, u_i, r_i, \bm{\varpi}_{i+1}\}$ is stored into the replay buffer $\mathcal{B}$. Thus, the mini-batch can be formed by randomly selecting  from the replay buffer. Third,  an exploration policy is applied to ensure adequate exploration of the continuous action spaces, where the action is selected from a random process that follows a Gaussian distribution with the mean $\pi(\bm{\varpi}_i;\theta_{\pi})$ and the variance $\varepsilon$. Here, $\varepsilon$ is an adjustable parameter to decay the randomness of the action selection.  Finally, the  paired pluggable modules  are shown in Fig. \ref{AN}.

\begin{figure*}[tbp]
  \centering
  \includegraphics[width=6in]{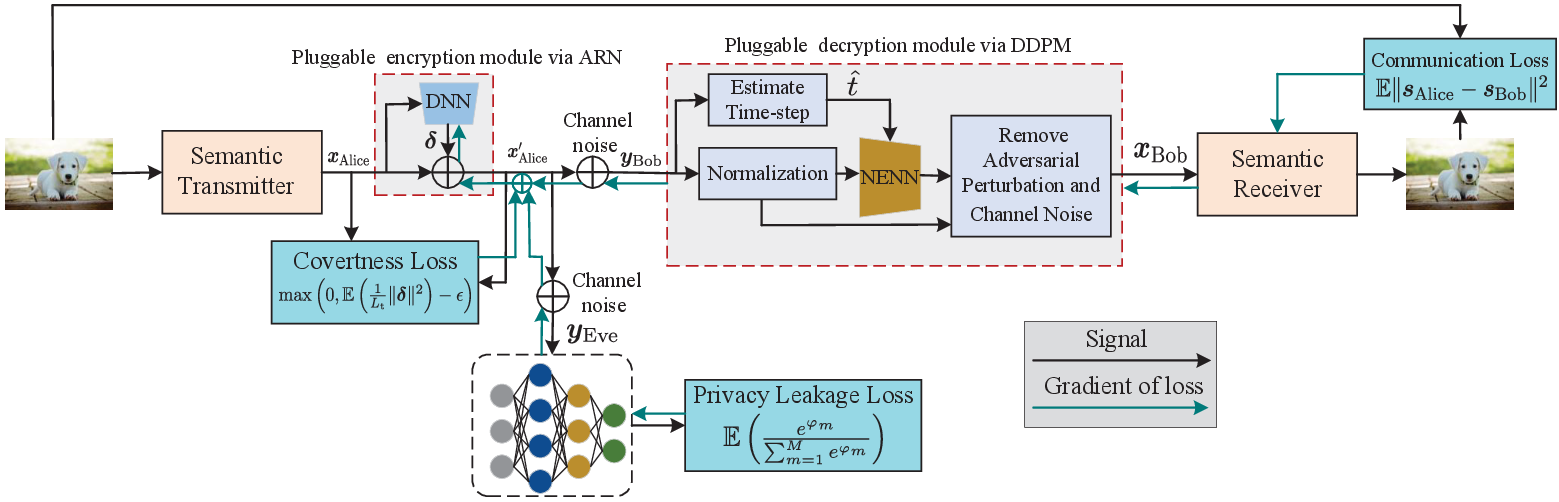}\\
  \caption{  Training procedure for ARN in the case of Alice having Eve's knowledge. Specifically,  the input data to DNN block consists of the output of the semantic transmitter. The stochastic gradient algorithm is used to minimize the loss function in problem \eqref{case2_SSC_problem_trans_replace_fur} to obtain the optimal ARN parameters. }  \label{ARN}
\end{figure*}

\subsection{Secure Paired Pluggable Modules Design for Alice having  Eve's knowledge}\label{has_eve_know}
In the previous section, we have discussed how to design $f_{\text{Alice}}$ and  $f_{\text{Bob}}$ when Alice has no the knowledge of Eve. In this section, we relax to the case where Alice has the knowledge of Eve, in which a tailored secure design for Eve can be performed. Without loss of generality, we consider the image label is the private information such as gender and item type, and Eve performs a classification task to steal Alice's private information.\footnote{Note that our design can be extended to other tasks directly, where only  the objective function for privacy leakage  is modified accordingly.  } Thus, the general optimization problem \eqref{SSC_problem} for the  security-aware semantic communication system can be further written as
\begin{align} \label{case2_SSC_problem_trans}
&\mathop{\text{min}}_{f_{\text{Alice}}, f_{\text{Bob}}}\lambda_\text{com}\mathbb{E}\|\bm{s}_{\text{Alice}}-\bm{s}_{\text{Bob}}\|^2-\lambda_\text{pri} {\text{Pr}\{g_{\text{Eve}}\left(\bm{y}_{\text{Eve}};{{\theta}_{\text{Eve}}}\right)\neq m\}}\notag\\
&~~~~~~~~~~~+\lambda_\text{cov} \mathbb{E}\|\bm{x}_{\text{Alice}}-\bm{x}_{\text{Alice}}'\|^2,
\end{align}
where $\text{Pr}\{\cdot\}$ is the probability function and $m$ denotes the true label for the image classification task. However, it is worth noting  that the probability function  has no explicit expression, and the optimization problem of maximizing the probability function is difficult. To alleviate this problem, a common approach is to use the surrogate  cross-entropy loss function $\mathcal{L}(\bm{y}_{\text{Eve}}, m; g_{\text{Eve}}, {\theta}_{\text{Eve}})$ instead of a probability function. Unfortunately, the value of   cross-entropy loss function may tend to infinity, making the training process extremely unstable. In this paper, the confidence of Eve's model for the true image label is used for the metric function of the privacy leakage. Specifically, the confidence is obtained using the softmax function, which is expressed by
\begin{align} \label{confi}
d_{\text{pri}}(\bm{s}_{\text{Alice}}, \bm{s}_{\text{Eve}})=\mathbb{E}\left(\frac{e^{\varphi_m}}{\sum_{m=1}^{M}e^{\varphi_m}}\right),
\end{align}
where $M$ is the  total number of image categories and
$\varphi_m$ is the output of the $m$-th neuron in  the final layer of Eve's classifier.  By replacing the probability function in \eqref{case2_SSC_problem_trans} with \eqref{confi}, we obtain
\begin{align} \label{case2_SSC_problem_trans_replace}
&\mathop{\text{min}}_{f_{\text{Alice}}, f_{\text{Bob}}}\:\lambda_\text{com}\mathbb{E}\|\bm{s}_{\text{Alice}}-\bm{s}_{\text{Bob}}\|^2+\lambda_\text{pri} \mathbb{E}\left(\frac{e^{\varphi_m}}{\sum_{m=1}^{M}e^{\varphi_m}}\right)\notag\\
&~~~~~~~~~~+\lambda_\text{cov} \mathbb{E}\|\bm{x}_{\text{Alice}}-\bm{x}_{\text{Alice}}'\|^2.
\end{align}

Next, our goal is to design $f_{\text{Alice}}$ and $f_{\text{Bob}}$ carefully according to the problem \eqref{case2_SSC_problem_trans_replace}. In the field of adversarial machine learning, numerous studies have shown that even small adversarial perturbations can significantly degrade the performance of deep learning models\cite{madry2017towards,moosavi2017universal}, which drives us to design the pluggable encryption module $f_{\text{Alice}}$ based on adversarial perturbations. The generated adversarial example at the pluggable encryption module $f_{\text{Alice}}$ is given by
\begin{align}
\bm{x}_{\text{Alice}}'&=f_{\text{Alice}}({\bm{x}}_{\text{Alice}};\theta_{\text{Alice}}) \label{f_alice_ap}\\
&=\underbrace{{\bm{x}_{\text{Alice}}}+\mathcal{F}\left({\bm{x}_{\text{Alice}}};\theta_{\text{Alice}}\right)}_{\text{Adversarial Residual Network}} \label{arn}\\
&={\bm{x}_{\text{Alice}}}+\bm{\delta},
\end{align}
where $\mathcal{F}(\cdot;\theta_{\text{Alice}})$ denotes a deep neural network (DNN) block; $\theta_{\text{Alice}}$ is the trainable parameters of DNN block; $\bm{\delta}$ is the adversarial perturbation generated by the DNN block; $\mathbb{E}\left(\frac{1}{{L_{\text{t}}}}\|\bm{\delta}\|^2\right)\leq \epsilon$ and $\epsilon$ is a small power limit. From the equation \eqref{arn}, we observe that \eqref{arn} essentially characterizes an ARN. Accordingly, the received signals of Bob and Eve are respectively written as
\begin{align}
\bm{y}_{\text{Bob}}&={\bm{x}_{\text{Alice}}}+\bm{\delta}+\bm{n}_{\text{Bob}}, \\
\bm{y}_{\text{Eve}}&={\bm{x}_{\text{Alice}}}+\bm{\delta}+\bm{n}_{\text{Eve}}.
\end{align}
Similar to the scenario where Alice lacks the knowledge of Eve, we use DDPM at Bob to generate the semantic information from the received noisy signal with  the adversarial perturbation and channel noise. Thus, in the case of Alice having Eve's knowledge, the  pluggable decryption module $f_{\text{Bob}}$ can be written by
\begin{align}
{\bm{x}}_{\text{Bob}}&= f_{\text{Bob}}\left(\bm{y}_{\text{Bob}};\theta_{\text{NENN}}^{*}\right)\label{f_bob_ap}\\
&=\frac{1}{\sqrt{\bar{\alpha}^{(\hat{t})}}}\left(\bar{\bm{y}}_{\text{Bob}}
-\sqrt{1-\bar{\alpha}^{(\hat{t})}}f_{\text{NENN}}\left(\bar{\bm{y}}_{\text{Bob}},\hat{t}; \theta_{\text{NENN}}^{*}\right)\right), \label{ap_denoise}
\end{align}
where $\bar{\bm{y}}_{\text{Bob}}$  is the normalized version of ${\bm{y}}_{\text{Bob}}$, i.e.
\begin{align}
\bar{\bm{y}}_{\text{Bob}}=\frac{{\bm{y}}_{\text{Bob}}}{\sqrt{1+\mathbb{E}\left(\frac{1}{L_{\text{t}}}\|\bm{\delta}\|^2\right)+\sigma^{2}_\text{b}}},
\end{align}
and  $\hat{t}$ is the estimated time-step, i.e.
\begin{align}
\hat{t}=\mathop{\text{argmin}}_{t}\left\|\frac{1}{\mathbb{E}\left(\frac{1}{L_{\text{t}}}\|\bm{\delta}\|^2\right)+\sigma^{2}_\text{b}}-\frac{\bar{\alpha}^{(t)}}{1-\bar{\alpha}^{(t)}}\right\|^2.
\end{align}

From \eqref{f_alice_ap} and \eqref{f_bob_ap}, the problem \eqref{case2_SSC_problem_trans_replace} can be further transformed to
\begin{align} \label{case2_SSC_problem_trans_replace_fur}
&\mathop{\text{min}}_{\theta_{\text{Alice}}}\:\lambda_\text{com}\underbrace{\mathbb{E}\|\bm{s}_{\text{Alice}}-\bm{s}_{\text{Bob}}\|^2}_{\text{Communication }}+\lambda_\text{pri} \underbrace{\mathbb{E}\left(\frac{e^{\varphi_m}}{\sum_{m=1}^{M}e^{\varphi_m}}\right)}_{\text{Privacy Leakage}}\notag\\
&+\lambda_\text{cov} \underbrace{\text{max}\left(0, \mathbb{E}\left(\frac{1}{{L_{\text{t}}}}\|\bm{\delta}\|^2\right)-\epsilon\right)}_{\text{Covertness}}.
\end{align}
The covertness loss function implies no penalty for portions below the power limit $\epsilon$, while severely penalizing portions that exceed the power limit $\epsilon$. It is worth noting that, to enable the trained ARN to adapt to different SNR environments, we model the noise power distributions of Bob and Eve as
\begin{align}
\sigma^2_{\text{Bob}} \sim \mathcal{U}(0, \Delta\sigma^2_{\text{Bob}}), \label{Bob_un}\\
\sigma^2_{\text{Eve}} \sim \mathcal{U}(0, \Delta\sigma^2_{\text{Eve}}),\label{Eve_un}
\end{align}
where $\mathcal{U}(a, b)$ denotes  the probability density function of a
random variable following the uniform distribution in the interval $[a,b]$; $\Delta\sigma^2_{\text{Bob}}$ and $\Delta\sigma^2_{\text{Eve}}$ are the maximal noise powers of interest. The input data to DNN block  consists of the output of the semantic transmitter. In each batch, the SNRs of Alice-Bob link and Alice-Eve link are static; however, they are re-sampled at the beginning of each new batch according to \eqref{Bob_un} and \eqref{Eve_un}. Once the optimal $\theta_{\text{Alice}}^{*}$ is obtained, the trained ARN can be deployed to generate the   adversarial perturbation for any output of the semantic transmitter. Finally, the training procedure of ARN is provided in Fig. \ref{ARN}.

\begin{remark}
 In Section \ref{SectionIV}, we have designed the paired pluggable modules for two scenarios: i) Alice lacks the prior knowledge of Eve, and ii) she has Eve's prior knowledge.. Specifically, in the absence of Eve's prior knowledge, the pluggable encryption module is designed to generate AGN. When Alice possesses prior knowledge of Eve, we use ARN to design the pluggable encryption module. In both scenarios, the pluggable decryption module is designed via DDPM.
Thus, when Alice transmits the private information, the paired pluggable modules can be plugged into the general semantic communication system against semantic eavesdropping. Conversely, when Alice transmits the public information, they can be unplugged.
\end{remark}

\section{Performance Analysis} \label{SectionV}
In this section, we empirically evaluate the performance and efficacy of our proposed scheme under diverse settings. In the following, we first present the experiment setup and then demonstrate  the performance of the proposed schemes.

\subsection{Experiment Setup}
In this paper, the proposed  diffusion-based scheme is evaluated on  multiple datasets including MNIST, CIFAR-10, and Fashion MNIST.  Due to space constraints, the primary experimental results are presented for the MNIST dataset, while the results for the other two datasets can be found in Section \ref{mul_data}. The   MNIST dataset  is   a large collection of handwritten digits commonly used for training and testing image processing systems.  The MNIST dataset contains $60,000$ training images and $10,000$ testing images, each of which is a $28\times28$ grayscale image of a single digit ranging from $0$ to $9$. The semantic encoder/decoder network and the joint source channel encoder/decoder network are fully connected neural networks with $32$ neurons each. Except for the semantic decoder, which uses the sigmoid function as its activation function, all other networks use the ReLU function as their activation function. The  size of the semantic encoder input and the channel input are respectively $28\times28=784$ and $23$. As shown in the equation \eqref{SC_MSE}, the MSE function is used for the loss function of the semantic communication system. The Adam optimizer is used to optimize the parameter set of the semantic communication system, where the learning rate is set to $0.001$ and the batch size is $256$. For the noise estimation neural network, a four-layer  fully connected neural network is used to estimate the noise, with each layer containing $128$ neurons. Each of the first three layers is followed by a time-step embedding layer, which makes the neural network obtain the time-step information of samples.  Furthermore, the first three layers all use the ReLU function as their activation function, while the fourth layer does not use any activation function. It is worth noting that the input and output sizes of the noise estimation neural network are identical, both equal to the size of the channel output, i.e., $23$. Similar to the training of the semantic communication system, the noise estimation neural network is trained using the Adam optimizer with a learning rate of $0.001$ and  a batch size of  $500$.

\begin{figure*}[tbp]
  \centering
  \includegraphics[width=5.5in]{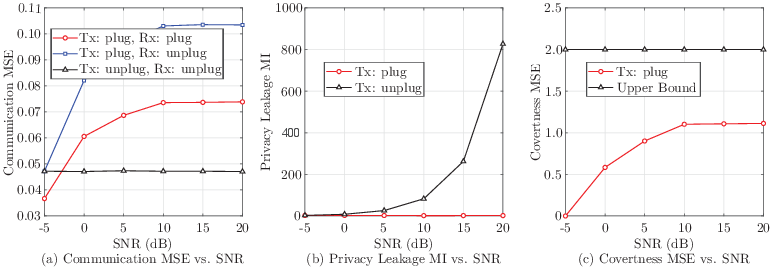}\\
  \caption{  The communication MSE, the privacy leakage MI, and the covertness MSE are evaluated with respect to (w.r.t.) the SNR of Alice-Eve, where the SNR of Alice-Bob link is fixed at $5$ dB. The results reveal that the designed paired pluggable modules  achieve excellent communication MSE while keeping the MI for privacy leakage close to zero.}  \label{AN_tradeoff}
\end{figure*}
For the actor network, a two-layer fully connected neural network is used to predict the action, where  each layer has $256$ neurons. The activation function for the first layer is ReLU function, and for the second layer, it is tanh function.  The input and output sizes of the actor network are respectively set to $2$ and $1$, which depend on the dimensions of  state space and  action space. Because the action values in this work are constrained to the range of $0$ to $1$, we scale the output $z$ of the second layer by applying the transformation ${(z+1)}/{2}$. For the critic network, a three-layer neural network with $256$ neurons in each layer, utilizing ReLU activation in the first and second layers,  is used to compute Q-values. The input and output sizes of the  critic network are $3$ and $1$, respectively.
The Adam optimizer with a learning rate of $0.01$ is used to optimize the actor network and the critic network. The soft update coefficient $\xi$ is $0.001$ and the Gaussian noise variance $\varepsilon$ is $0.5$. The hyperparameters $\lambda_{\text{com}}$, $\lambda_{\text{pri}}$, and $\lambda_{\text{per}}$ are set to $10$, $0.25$, and $1$, respectively.
Note that, unless otherwise specified, the above experiment setup are used for the  default parameters.

\subsection{Performance Evaluation for Unknown Eve's Knowledge}

\noindent\emph{\textbf{Observation 1:} The designed pluggable encryption module via AGN achieves a  near-zero level of privacy leakage MI, while the designed paired pluggable decryption module via DDPM generates the detailed semantic information from the received noisy signal with the actively introduced AGN and the passively introduced channel noise well, and thus achieve a superior semantic communication MSE on the legitimate communication
link.
(cf. Fig. \ref{AN_tradeoff})}

In Fig. \ref{AN_tradeoff}, we plot the communication MSE, the privacy leakage MI, and the covertness MSE as functions of SNR of Alice-Eve link, where the SNR of Alice-Bob link is fixed at $5$ dB and the SNR of Alice-Eve link varies from $-5$ dB to $20$ dB. {\textbf{``Tx: plug, Rx: plug" means inserting a pluggable encryption module after the semantic transmitter and a pluggable decryption module before the semantic receiver. Similarly, the meanings of other legends can be obtained.}} ``Upper Bound" means that all transmission power is used to send AGN, which corresponds to the upper bound of the covertness MSE, i.e., $2$.
From Fig. \ref{AN_tradeoff}(a), it can be seen that the designed pluggable decryption module via DDPM achieves significantly better communication MSE compared to when it is not inserted. This is mainly because that the designed pluggable decryption module can well generate  the semantic information from the received noisy signal with AGN and channel noise. Furthermore, at $\mathsf{SNR}=-5$ dB, the pluggable decryption module makes the communication MSE even better than that of the original semantic communication system (Tx: unplug, Rx: unplug). From
Fig. \ref{AN_tradeoff}(b), we observe that: If the pluggable module is not inserted at the transmitter, the MI for privacy leakage on the Alice-Eve link will approach infinity as the SNR increases. Conversely, inserting the pluggable module after the semantic transmitter ensures that the MI for privacy leakage remains at a low level (around $2$).  In
Fig. \ref{AN_tradeoff}(c),  We can notice from Fig. \ref{AN_tradeoff}(c) that as the SNR increases, although the covertness MSE also increases, it does not exceed $1.2$ and remains well below its upper bound. The low covertness MSE means that the channel input has only a slight distortion.

\begin{figure}[tbp]
  \centering
  \includegraphics[width=3in]{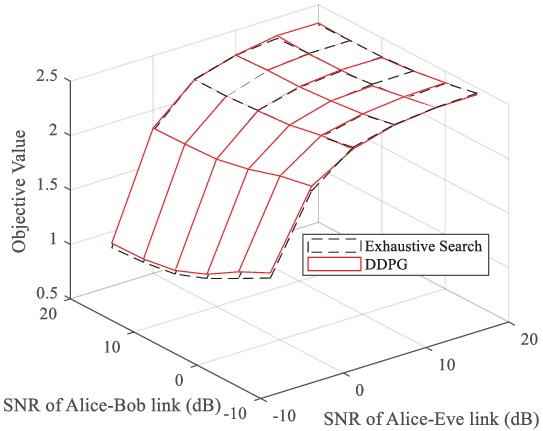}\\
  \caption{  The objective  value obtained by the exhaustive search method and the proposed DDPG scheme are evaluated w.r.t.  SNRs of the Alice-Bob and Alice-Eve links. The results indicate that the proposed DDPG method achieves a performance close to that of the exhaustive search method.}  \label{actor_reward}
\end{figure}

\noindent\emph{\textbf{Observation 2:} The trained actor network is capable of quickly determining the $u$ value while achieving a performance close to that of the exhaustive search method.
(cf. Fig. \ref{actor_reward})}

\begin{figure*}[tbp]
  \centering
  \includegraphics[width=5.3in]{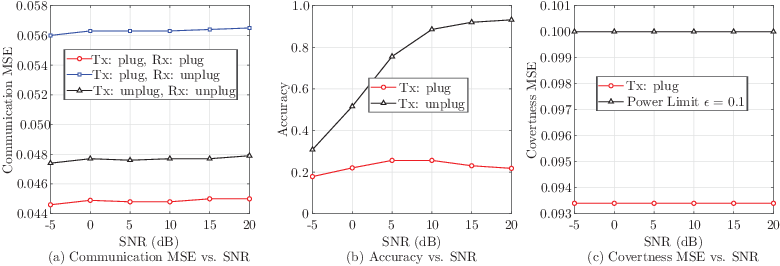}\\
  \caption{  The communication MSE, the Eve's classification accuracy, and  covertness MSE are evaluated w.r.t. the SNR of Alice-Eve link, where the SNR of Alice-Bob link is $5$ dB. It is worth noting that the covertness MSE corresponds to the power of the adversarial perturbations. The results show the designed pluggable encryption module via ARN can   generate low-power perturbation signals that successfully mislead Eve's deep learning model with a high probability. Additionally, the pluggable decryption module via DDPM  effectively eliminates these perturbations and channel noise, thereby achieving the high-quality semantic communication.}  \label{ARN_MSE_ACC_power}
\end{figure*}

We compare the objective values obtained by the exhaustive search method and the proposed DDPG scheme under different SNR environments, as depicted in Fig. \ref{actor_reward}, where the objective value denotes the weighted sum of  the communication MSE, the privacy leakage MI, and the covertness
MSE, as shown in \eqref{SSC_problem_trans_u_MI}. The step size for the exhaustive search method is set to $0.01$, which indicates that $100$ searches are required. First, we can see that, in different SNR environments, the proposed DDPG scheme achieves performance close to that of the exhaustive search method. Second, we compared the running times of the DDPG and exhaustive methods. The results show that the actor network  predicts the $u$ value based on the current environment in only $0.1492$ seconds, whereas the exhaustive method requires $18.68$ seconds. Third, it can be observed that, as the channel quality of the Alice-Eve link deteriorates, the objective value gradually decreases to below $1$. This is primarily because, when the SNR of the Alice-Eve link is low, the pluggable encryption module only needs to use the low-power AGN to ensure a small privacy leakage MI.

\subsection{Performance Evaluation for Known  Eve's Knowledge}

\begin{figure}[tbp]
  \centering
  \includegraphics[width=3in]{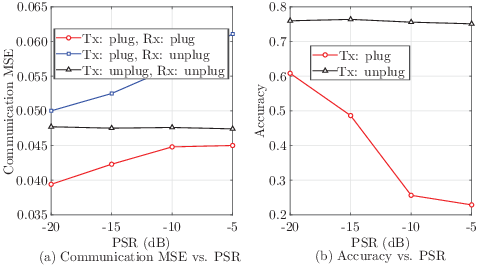}\\
  \caption{  The communication MSE and the Eve's classification accuracy are evaluated w.r.t.  PSR, where the SNR of Alice-Bob and Alice-Eve links is $5$ dB. The results indicate that, with the designed pluggable encryption and decryption modules, increasing the perturbation power further reduces Eve's classification accuracy while still achieving an acceptable  MSE.}  \label{ARN_MSE_ACC_vs_power}
\end{figure}

\begin{figure}[tbp]
  \centering
  \includegraphics[width=2.1in]{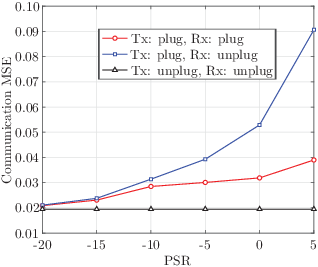}\\
  \caption{  The communication MSE is evaluated w.r.t. the PSR, where the channel noise of Alice-Bob link is set to $0$. The results reveal that the pluggable decryption module via DDPM is capable of removing  the adversarial perturbations.}  \label{ARN_no_noise}
\end{figure}

\noindent\emph{\textbf{Observation 3:} The pluggable encryption module via ARN  reduces Eve's classification accuracy to a  near-failure level while the pluggable decryption module via DDPM achieves a communication MSE comparable to that of the original semantic communication system.
(cf. Figs. \ref{ARN_MSE_ACC_power}, \ref{ARN_MSE_ACC_vs_power}, and \ref{ARN_no_noise})}

In Fig. \ref{ARN_MSE_ACC_power}, we evaluate the communication MSE, Eve's classification accuracy, and the covertness MSE w.r.t. the SNR of Alice-Eve link when Alice has the knowledge of Eve, where the SNR of Alice-Bob link is $5$ dB and that of  Alice-Eve link varies from $-5$ dB to $20$ dB. Here,  the covertness MSE denotes the power of the adversarial perturbations. From Fig. \ref{ARN_MSE_ACC_power}(a), we observe that  inserting the pluggable encryption module increases the communication MSE, but inserting the pluggable decryption module reduces the communication MSE to a level lower than that of the original semantic communication system (Tx: unplug, Rx: unplug).
This is because the designed pluggable decryption module based on DDPM can generate the semantic
information from the received noisy signal with the adversarial perturbations  and the channel noise.
Fig. \ref{ARN_MSE_ACC_power}(b) compares the impact of inserting and removing the pluggable modules on Eve's classification accuracy. Clearly, inserting the pluggable encryption module reduces Eve's classification accuracy to around $0.2$ (say, a near-failure level). This means that the adversarial samples generated by ARN can prevent Eve from obtaining the private label information. From Fig. \ref{ARN_MSE_ACC_power}(c), we can see that the  power of the adversarial perturbation remains consistently below the power limit of $0.1$. Notably, the power of the original semantic transmitter's output $\bm{x}_{\text{Alice}}$ is $1$, which means the power of the adversarial perturbation is less than $10\%$ of the power of $\bm{x}_{\text{Alice}}$. The above observations indicate that, in the case of having  Eve's knowledge, the designed pluggable  encryption and decryption modules can achieve high-quality semantic communication while misleading Eve's image classifier and maintaining high covertness.

We evaluate the communication MSE and the classification accuracy w.r.t. the perturbation-to-signal ratio (PSR) when Alice has the knowledge of Eve, as shown in Fig. \ref{ARN_MSE_ACC_vs_power}, where PSR denotes the power ratio of the adversarial perturbation and the output $\bm{x}_{\text{Alice}}$ of the semantic transmitter. Note that  the SNR of both the Alice-Bob and Alice-Eve links is $5$ dB.
As shown in  Fig. \ref{ARN_MSE_ACC_vs_power}(a), the communication MSE increases with the PSR. However, installing a pluggable decryption module before the semantic receiver significantly reduces the communication MSE, even surpassing the performance of the original semantic communication system (Tx: unplug, Rx: unplug). From Fig. \ref{ARN_MSE_ACC_vs_power}(b), we observe that an increase in PSR  reduces Eve's classification accuracy. When the PSR reaches $-5$ dB, Eve's classification accuracy is almost on the verge of failure, while the communication MSE remains around $0.045$.

\begin{figure}[tbp]
  \centering
  \includegraphics[width=2.13in]{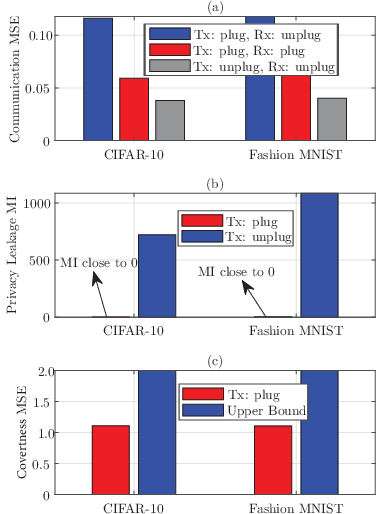}\\
  \caption{ The communication MSE, the privacy leakage MI, and  the covertness MSE are evaluated on CIFAR-$10$ and Fashion MNIST datastes when Alice has no the Eve's knowledge, where the SNR of the Alice-Bob link is  $5$ dB and that of the Alice-Eve link is  $10$ dB. (a) Communication MSE; (b) Privacy leakage MI; (c) Covertness MSE. The results indicate that the designed paired pluggable modules well balance the communication MSE and the privacy leakage MI.}  \label{unknow}
\end{figure}

In Fig. \ref{ARN_no_noise}, we  evaluate the communication MSE as a function of PSR. Note that the channel noise for the Alice-Bob link is set to $0$, indicating the presence of only adversarial perturbations. This is done to assess the capability of the pluggable decryption module via DDPM in mitigating adversarial perturbations.  From Fig. \ref{ARN_no_noise}, it can be seen that as PSR increases, the communication MSE continuously increases due to the pluggable encryption module via ARN. When the PSR is $5$ dB, the communication MSE reaches $0.09$, while the MSE of the original semantic communication system remains around $0.02$. Encouragingly, by inserting the pluggable decryption module via DDPM, the communication MSE shows only a slight increase compared to the original semantic communication system. This result indicates that the pluggable decryption module via DDPM well generate the detailed semantic
information from the received noisy signal with  adversarial perturbations.

\subsection{Performance Evaluation On Multiple Datasets} \label{mul_data}

\begin{figure}[tbp]
  \centering
  \includegraphics[width=2.08in]{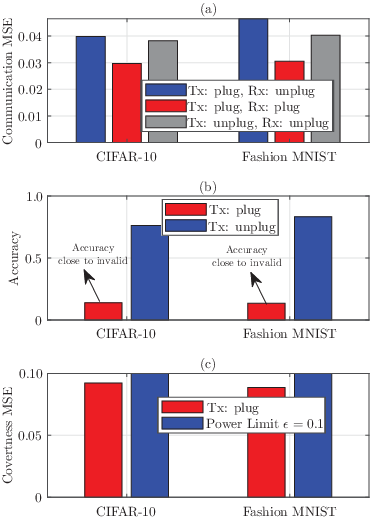}\\
  \caption{  The communication MSE, the Eve's classification accuracy, and  the covertness MSE are evaluated on CIFAR-$10$ and Fashion MNIST datastes when Alice has  the Eve's knowledge, where the SNRs of the Alice-Bob link and the Alice-Eve link are  $5$ dB and $10$ dB, respectively.
  (a) Communication MSE; (b) Eve's classification accuracy; (c) Covertness MSE.  The results show that the designed paired pluggable modules well balance the communication MSE and the Eve's classification accuracy.}  \label{know}
\end{figure}

\noindent\emph{\textbf{Observation 4:} On multiple different datasets (say, CIFAR-$10$ and Fashion MNIST), the designed pluggable modules successfully prevent the semantic eavesdropping while ensuring the high-quality semantic communication.
(cf. Figs. \ref{unknow} and \ref{know})}

We evaluate the communication MSE, privacy leakage MI, and  covertness MSE on CIFAR-$10$ and Fashion MNIST datastes when Alice has no the Eve's knowledge, as depicted in Fig. \ref{unknow}, where the SNR of the Alice-Bob link is set to $5$ dB and that of the Alice-Eve link is  $10$ dB. Note that similar conclusions can be obtained under other SNR configurations. For  brevity, we do not present these results here. Different  from the MNIST dataset, the CIFAR-$10$ dataset consists of $60,000$ $3\times32\times32$ color images in $10$ different classes, with $6,000$ images per class. The $10$ classes include airplanes, automobiles, birds, cats, deer, dogs, frogs, horses, ships, and trucks. Fashion MNIST is a dataset designed as a more challenging replacement for the original MNIST dataset. Fashion MNIST comprises $70,000$ grayscale images, each $28\times28$ pixels in size, divided into $10$ classes representing different types of clothing items. The $10$ classes are t-shirt, trouser, pullover, dress, coat, sandal, shirt, sneaker, bag, and ankle boot. From Fig. \ref{unknow}, we can observe that the AGN-based pluggable encryption module reduces the MI leakage of privacy information to near zero, indicating that it can effectively prevent Eve's semantic eavesdropping. At the same time, the DDPM-based pluggable decryption
module ensures that the communication MSE only has a slight increase compared to the original semantic communication system. In addition, we can see that the channel input has a slight distortion due to the  low covertness MSE.

In Fig. \ref{know}, we evaluate the communication MSE, Eve's classification accuracy, and  covertness MSE on CIFAR-$10$ and Fashion MNIST datastes when Alice has   Eve's knowledge, where the SNRs of the Alice-Bob and Alice-Eve links are  $5$ dB and $10$ dB, respectively. As shown in Fig. \ref{know}, the ARN-based pluggable encryption module reduces Eve's classification accuracy to a near-failure level, meaning that Eve obtains the semantic information of  label with a high probability of error. However, by using the DDPM-based pluggable decryption module, the communication MSE is even better than that of the original semantic communication system. Encouragingly, when Alice has the knowledge of Eve, the covertness MSE can be controlled to a low level (say, below $0.1$).
\section{Conclusion} \label{SectionVI}
In this paper, we proposed a diffusion-enabled semantic communication system against eavesdropping, which can be applied to  both scenarios where Alice has or does not have the knowledge of Eve.  Specifically,  we proposed designing AN including type and power to prevent eavesdropping and explored the reverse process of diffusion models to adaptively generate the detailed semantic information from the received noisy signal with the actively
introduced AN and the passively introduced channel noise. We developed the  paired pluggable encryption and decryption modules tailored for the semantic communication systems. Simulation results show that the designed pluggable modules successfully prevent semantic eavesdropping while achieving the high-quality semantic communication. Additionally, the generated AN causes only slight distortion of the channel input (say, covertness). Furthermore,
the designed pluggable modules  avoid  retraining the semantic communication system, thereby saving significant computational overhead.

 Different from traditional literature, we  first carefully design AN  (say, Gaussian noise or adversarial perturbation) in terms of type and power to counteract semantic eavesdropping, and then map the AN and  channel noise to the forward diffusion process, and finally propose the adaptive inverse process to  generate the semantic information from the noisy signal with the AN  and the natural noise. Here, the power of AGN is obtained using DDPG, and we  explore the positive side of adversarial perturbations to prevent eavesdropping. Ultimately, the designed scheme achieves a good balance between the quality of service  and the privacy protection.
Excitingly, even using simple fully connected neural networks as the noise estimation network for DDPM can effectively generate the semantic information from noisy data. To further enhance the performance of the diffusion-enabled pluggable modules, future work is expected to incorporate more advanced GAI techniques, such as transformers. Additionally, conducting performance evaluations on a broader range of datasets will be interesting. Finally, leveraging transfer learning and other machine learning methods to extend task-specific pluggable modules into task-agnostic ones is also a promising direction.

\bibliographystyle{IEEEtran}
\bibliography{arxiv_DIFFUSION_submit}

\end{document}